# Competing oxygen evolution reaction mechanisms revealed by high-speed compressive Raman imaging

[1,2]Raj Pandya[✉], [3,4]Florian Dorchies[#], [5,6]Davide Romanin[#], [1]Sylvain Gigan, [5]Alex W. Chin, [1]Hilton B. de Aguiar[✉] and [3,4,7]Alexis Grimaud[✉]

[1]Laboratoire Kastler Brossel, ENS-Université PSL, CNRS, Sorbonne Université, Collège de France, 24 rue Lhomond, 75005 Paris, France

[2]Department of Physics, Cavendish Laboratory, University of Cambridge, JJ Thomson Avenue, Cambridge CB3 0HE, United Kingdom

[3]Chimie du Solide et de l'Energie, UMR 8260, Collège de France, Paris, France

[4]Réseau sur le stockage Electrochimique de l'Energie (RS2E), Amiens, France

[5]Sorbonne Université, CNRS, Institut des Nanosciences de Paris, UMR7588, F-75252, Paris, France

[6]Université Paris-Saclay, CNRS, Centre de Nanosciences et de Nanotechnologies, 91120, Palaiseau, France

[7]Department of Chemistry, Boston College, Merkert Chemistry Center, 2609 Beacon St., Chestnut Hill, MA, 02467 USA

[✉]Correspondence to: rp558@cam.ac.uk, h.aguiar@lkb.ens.fr, alexis.grimaud@bc.edu

[#]These authors contributed equally

**Abstract**

**Transition metal oxides are state-of-the-art materials for catalysing the oxygen evolution reaction (OER), whose slow kinetics currently limit the efficiency of water electrolysis. However, microscale physicochemical heterogeneity between particles, dynamic reactions both in the bulk and at the surface, and an interplay between particle reactivity**

**and electrolyte makes probing the OER challenging. Here, we overcome these limitations by applying state-of-the-art compressive Raman imaging to uncover competing bias-dependent mechanisms for the OER in a solid electrocatalyst, α-$Li_2IrO_3$. By spatially and temporally tracking changes in the in- and out-of-plane Ir-O stretching modes – identified by density functional theory calculations – we follow catalytic activation and charge accumulation following ion exchange under a variety of electrolytes, particle compositions and cycling conditions. We extract velocities of phase fronts and demonstrate that at low overpotentials oxygen is evolved by the combination of an electrochemical-chemical mechanism and a classical electrocatalytic adsorbate mechanism, whereas at high overpotentials only the latter occurs. These results provide strategies to promote mechanisms for enhanced OER performances, and highlight the power of compressive Raman imaging for low-cost, chemically specific tracking of microscale reaction dynamics in a broad range of systems where ion and electron exchange can be coupled to structural changes, *i.e.* catalysts, battery materials, memristors, *etc*.**

Introduction

The oxygen evolution reaction (OER) is the anodic half-reaction generating protons and electrons necessary for the electrochemical synthesis, in aqueous conditions, of chemical fuels such as hydrogen or $CO_2$-reduction products (CO, formic acid, ethylene *etc*). Boosting OER kinetics – a reaction involving the exchange of four electrons and four protons which can be more sluggish than its cathodic counterparts – is critical to ensure good performance for water or $CO_2$ electrolyzers[1,2]. In pursuit of that goal, recent years have seen a surge in physical and chemical descriptors helping the design of enhanced OER catalysts, often focussing on the state-of-charge of the catalysts independent of the applied bias[3–9].

Recent studies have however challenged this view by demonstrating for several iridium oxide catalysts that the accumulation of charges upon OER[10–12] directly impacts their electrochemical response. Compared to the more classical case where the charge of the catalyst is constant, oxides storing charges are currently regarded as having higher activity and stability in both acidic and alkaline conditions, making them a promising new class of catalysts. In acidic conditions, amorphous[10] $IrO_x$ compounds have been shown to be oxidized, *i.e.* deprotonated, to accumulate *O species. The catalytic rate, *i.e.* the OER current, then depends linearly on the accumulated charge, with the *O species consumed when releasing the bias[10]. Similarly, it was shown in crystalline layered iridium oxides that there is a build-up of charge upon bulk reversible exchange of cations, protons in acidic conditions[11] or hydrated potassium cations in alkaline conditions[13]. Specifically, in these layered iridium oxides, bulk oxidation/charging is followed by the chemical reaction of the oxidized form of the catalyst with water. This results in the evolution of oxygen, while charge balance of the catalyst is ensured by bulk intercalation of four cations per molecules of oxygen. The oxidized form of the catalyst is then regenerated by electrochemical deintercalation, thus accumulating charges back and encompassing a so-called Electrochemical-Chemical (EC) mechanism as shown in **Figure 1a**[13].

Very little is known regarding the dynamics of this bulk Electrochemical-Chemical process, directly limiting our ability to master this mechanism and further improve its kinetics to develop better OER catalysts. This is due to the lack of characterization techniques with sufficient time-, space- and chemical-resolution to quantitatively measure the extent to which cations exchange upon OER[14]. Indeed, for catalysts storing charges, a release of potential in the presence of electrolyte will quench the charge, meaning *ex-situ* methods cannot be used to capture the OER dynamics or even the real active state of the catalyst. One piece of information that is critically missing is a quantitative evaluation of the speed at which phase fronts propagate. If the velocity of intercalating cations is much slower than the turnover frequency

(TOF) for the OER on surface active sites, the EC mechanism would be limited to the surface extremities. In this case the EC mechanism would mainly serve to quench the accumulated charges upon rest and stabilize the surface, with the OER proceeding through a classical electrochemically-driven proton couple mechanism (**Figure 1a** left panel). In contrary, if the velocity of cations is high, intercalation may proceed through a non-negligible portion of the particle making the EC mechanism an important contributing pathway for molecular oxygen generation.

Spontaneous Raman spectroscopy, and surface enhanced variants[15,16], are widely used for tracking of electrode dynamics with chemical bond-selectivity[17]. Indeed, a number of previous works have used Raman spectroscopy specifically for *operando* investigation of the OER mechanism[18–21]. However, the low-signals in spontaneous Raman have hampered obtaining the necessary spatial and temporal-resolution for real-time Raman imaging of operating electrodes. Furthermore, many electrochemical systems demand probing in the 300 to 1200 cm$^{-1}$ spectral region, which is challenging for the more sensitive coherent Raman variants[22]. The data loads and relative expense in-terms of equipment[23], as compared to other optical methods, *e.g.* reflection microscopy[24,25], have additionally limited the use of Raman as a laboratory tool for imaging inside electrochemical systems. In this work, we alleviate these limitations by using the compressive Raman imaging scheme[26–28] to demonstrate how high-resolution spontaneous Raman imaging can elucidate the charge compensation mechanism in charge-accumulating OER catalysts. Tracking changes in the in- and out-of-plane vibrational stretches of a layered α-Li$_2$IrO$_3$ catalyst we follow the dynamics of cation intercalation upon cycling, as well as its potential dependence. Importantly our high temporal and spatial resolution allows us to extract the velocity for different phase fronts inside the catalyst, thus providing a definitive answer regarding the predominant OER mechanism at play on the surface of this novel class of catalysts.

**Main**

**Figure 1b** shows a schematic of the spontaneous Raman imaging setup. Excitation of the samples is performed with a 532 nm laser which is focussed onto the sample using a 1.4 numerical aperture objective (spatial resolution ~300 nm). The laser beam is rapidly raster scanned across the sample using galvanometric mirrors with the inelastic Raman backscatter light guided to spectrometers with a confocal pinhole to ensure imaging from a single plane. For measurements which are only spectrally resolved *i.e.* no imaging, we use a 'conventional' spectrometer equipped with an electron-multiplying charge-coupled (EMCCD) camera to acquire the whole vibrational spectrum in a single-shot. For Raman imaging requiring higher readout and reduced data loads we exploit the compressive Raman imaging method. Here, a programmable spectrometer equipped with a single-pixel detector – in our case a single-photon avalanche photodiode (SPAD) – is used[29]. Such a framework allows us to significantly compress the spectral data during the measurement stage and remove the excess noise of conventional cameras when imaging at high-speed[23]. As the beam is raster-scanned across the sample, optimal filters loaded into the programmable spectrometer allow selecting of the Raman scattered light associated with each of the predetermined species of interest and background (see **Supplementary Note 1** for further details). Orthogonality between filters, whose widths are sufficiently large enough to account for any spectral shifts, ensures no overlap between spectral channels. In this way chemical imaging can be achieved at up to ~0.3 fps (50 μm²). A maximum laser power of ~50 mW (before objective) is used in all experiments to minimise sample degradation (see **Supplementary Note 2**).

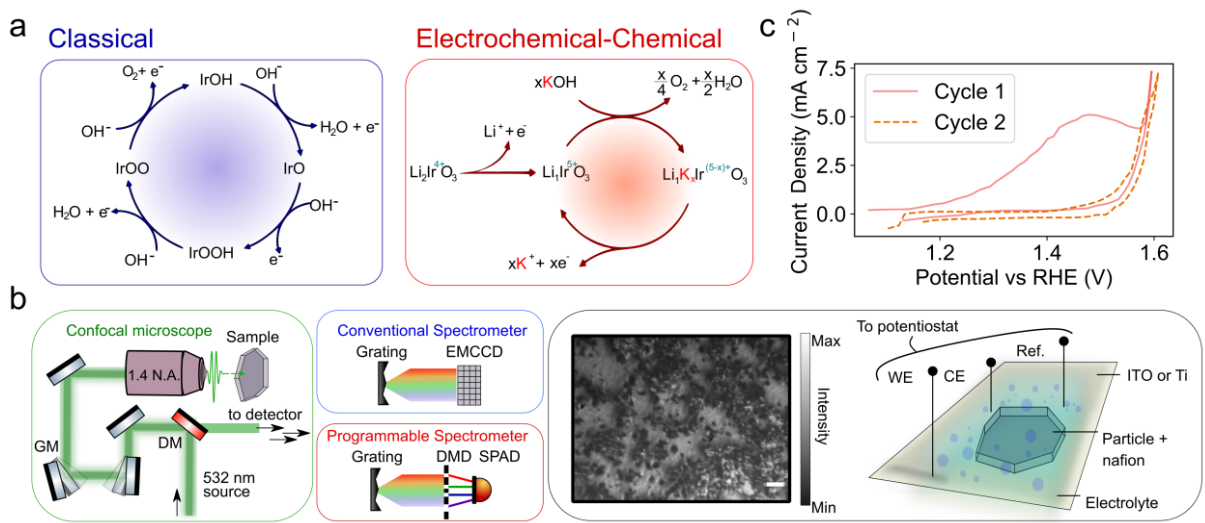

**Figure 1: a.** Classical "adsorbate" mechanism for water oxidation on the surface of iridium oxide catalysts (blue box) and electrochemical-chemical mechanism (red box) for water oxidation with α-Li$_2$IrO$_3$. **b.** (Green box) Cartoon schematic of confocal Raman microscope. The 532 nm laser is scanned across a sample using galvanometric mirrors (GM) with the excitation and Raman light separated by a dichroic mirror (DM). Either a conventional spectrometer (equipped with EMCCD (conventional detector; blue box)) or a programmable spectrometer (equipped with a DMD (digital micromirror device) and avalanche photodiode (SPAD; red box)) are used to measure the Raman signal. (Black box) Brightfield image of α-Li$_2$IrO$_3$ particles dispersed with Nafion onto ITO slide (left). Scale bar is 10 μm. (Right) Schematic of three-electrode setup for *operando* electrochemical cycling. Light is focussed onto the sample through the back side of an ITO or Ti coated coverslip which acts as the working electrode (WE). **c.** Cyclic voltammogram of α-Li$_2$IrO$_3$ showing the first two cycles. Scan rate is 10 mV/s, current is normalized by the geometric surface area.

α-Li$_2$IrO$_3$ particles are dispersed onto thin coverslides with a conductive coating (ITO or 10 nm of Ti, depending on exact experiment, with negligible difference in cycling/Raman response between substrates, see **Supplementary Note 2**; **Figure 1b**), with potentials applied using a standard three-electrode setup consisting of a Pt wire counter electrode and Ag/AgCl reference electrode. A 1M KOH solution is used as the electrolyte. As is often observed for layered intercalation compounds synthesized by high temperature solid-state routes, platelet-like α-Li$_2$IrO$_3$ particles show a preferential orientation after deposition with (00$l$) facets being exposed (as schematically represented in **Figure 1b**). All imaging in this manuscript is hence along this crystallographic direction and performed on 'flat' particles (see **Experimental Methods**). Similar electrochemical behaviour is observed with this setup, as shown in **Figure 1c**, as compared to that recorded using a classical three-electrode cell with the catalyst particles drop-casted onto a rotating disk electrode as done previously[13]. A first anodic activation is observed (current peak centred at ~1.45 V vs the reversible hydrogen electrode (RHE)), which has been associated with the deintercalation of $Li^+$ from α-$Li_2IrO_3$ to a stoichiometry close to α-$Li_1IrO_3$; we refer to this compound as the activated phase in the following. At more anodic potentials, OER current is measured. Once formed, the oxidized form of the catalyst α-$Li_1IrO_3$ can chemically react with KOH following the EC mechanism described in **Figure 1a**, as we previously demonstrated[13]. This reaction leads to the intercalation of hydrated potassium cations in the bulk of the catalyst to form a birnessite phase and the evolution of molecular oxygen; we refer to this latter compound as α-$Li_1K_xIrO_3$ henceforth. Nevertheless, the extent to which this mechanism occurs and its exact potential range of operation remains unknown due to our inability to track the reaction with sufficient simultaneous temporal, spatial and chemical-resolution. After the initial anodic scan, no signs of irreversible oxidation are observed and, based on electrochemical analysis, hydrated $K^+$ was proposed to be reversibly exchanged at potentials concomitant to the OER[13].

Using the aforementioned setup, *ex-situ* Raman spectra were collected for α-$Li_2IrO_3$, *i.e.* before cycling (pristine), after the initial activation/delithiation and after cycling during which potassium is intercalated to form α-$Li_1K_xIrO_3$ (7 cycles; **Supplementary Note 3**). As shown in **Figure 2a**, the Raman spectrum of α-$Li_2IrO_3$ shows two peaks at ~550 $cm^{-1}$ and ~640 $cm^{-1}$, in agreement with previous studies[30,31]. Activation/delithiation induces a modification of the Raman spectrum for the catalyst with a ~50% dimming of the 640 $cm^{-1}$ mode and ~20% dimming of the ~550 $cm^{-1}$ mode with a red shift of the peak centre-of-mass (CoM) of ~5 to 15 $cm^{-1}$ (absolute peak position resolution is ~12 $cm^{-1}$, hence CoM is instead reported; see **Supplementary Note 3**). On cycling to form α-$Li_1K_xIrO_3$, the Raman mode at 640 $cm^{-1}$ grows in intensity, but only to ~40% of the intensity of that in α-$Li_2IrO$ (**Figure 2a**). The 550 $cm^{-1}$ mode intensity and CoM remains relatively unperturbed as compared to the activated form.

Density functional theory (DFT) calculations were carried out to unravel the atomic displacements associated with the two observed Raman peaks for both the pristine α-$Li_2IrO_3$ compound and the catalyst containing potassium after reaction with KOH (**Supplementary Note 4**). For both compounds the mode at ~550 $cm^{-1}$ corresponds to stretching/scissoring of the Ir-O bond in the plane of $IrO_6$ edge-sharing octahedra layers, whereas the higher frequency mode at ~640 $cm^{-1}$ corresponds to an Ir-O stretch out of the layer plane. We note that on potassium intercalation the computed in-plane 550 $cm^{-1}$ mode shows relatively little change in spectral position whereas at higher frequencies around 640 $cm^{-1}$ the computed shifts are larger (~50 $cm^{-1}$ blueshift). In general, the larger changes in the 640 $cm^{-1}$ mode as compared to that at 550 $cm^{-1}$ can be rationalised by realising that the former involves modifications along the *c*-direction which are much stronger (increase of more than 1 Å for the interlayer distance upon hydrated $K^+$ intercalation, as previously probed by XRD[13]) when compared to those in the *ab* plane. Although the comparison between DFT and experiment is qualitative, the agreement between them is strong. Furthermore, both experiment and theory are consistent with previous studies of layered intercalation compounds where modes of $A_{1g}$-like symmetry (in-plane, ~550 $cm^{-1}$) and $E_g$-like symmetry (out-of-plane, ~640 $cm^{-1}$) drop in intensity on deintercalation[32].

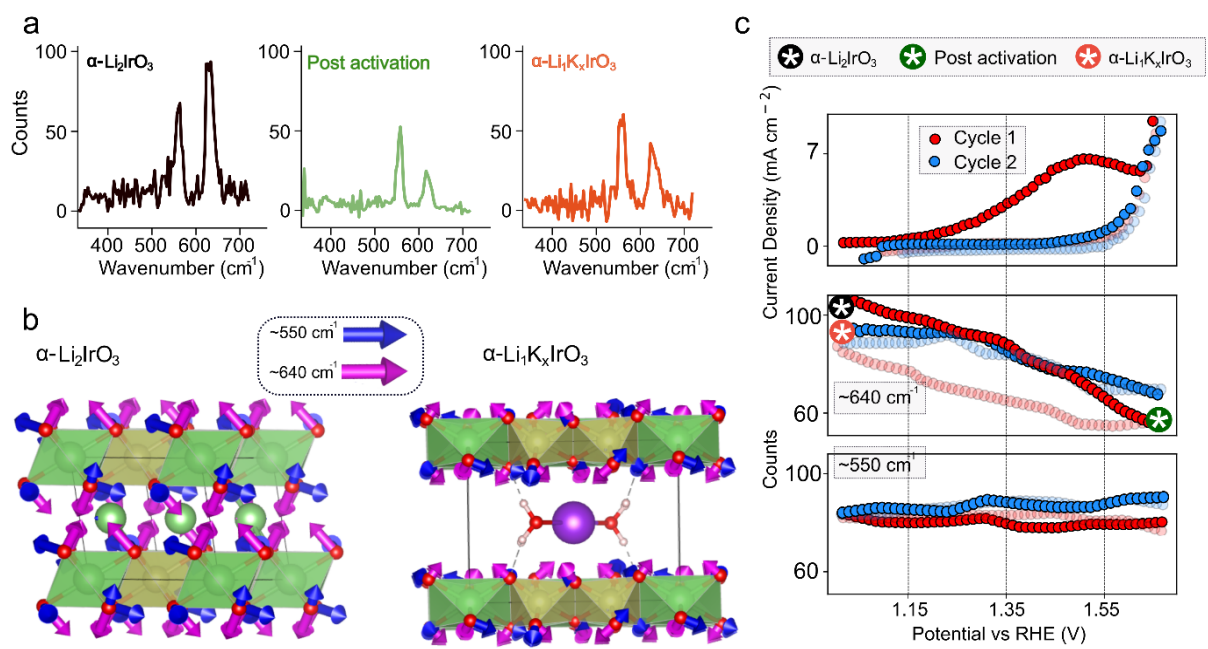

**Figure 2: a.** Raman spectra experimentally recorded for α-Li$_2$IrO$_3$, post oxidation/delithiation ('post activation') and α-Li$_1$K$_x$IrO$_3$. **b.** Crystallographic structures of α-Li$_2$IrO$_3$ and α-Li$_1$K$_x$IrO$_3$ with arrows highlighting computed atomic motion corresponding to Raman modes at ~550 cm$^{-1}$ and 640 cm$^{-1}$. **c.** Cyclic voltammograms recorded over the first two cycles of α-Li$_2$IrO$_3$ with corresponding intensity of the 640 cm$^{-1}$ and 550 cm$^{-1}$ modes. Cycling was performed at 10 mV/s due to acquisition speed limitations of Raman setup. First and second cycles are shown in red and blue, respectively, with dark circles showing anodic sweep, and faded circles the cathodic one. Coloured circles with white asterisk mark phases formed at respective start or end-points of a given cycle.

To further understand the link between cation exchange and changes in the Raman spectrum, *operando* measurements using the conventional spectrometer are carried out with spectra measured during cyclic voltammetry in 1M KOH (aq) electrolyte (**Figure 2c**). First, we remark that the Raman experiments in **Figure 2** and **Figure 3** below are done at scan rates of 10 mV/s (non-spatially resolved Raman measurements) and 4 mV/s (spatially resolved measurements) to allow for effective synchronisation with our detectors. During the initial anodic scan (1.1 V to 1.7 V vs RHE), an immediate dimming of the 640 cm$^{-1}$ (ΔCounts ~40%) and to a much lesser extent the 550 cm$^{-1}$ (ΔCounts ~5%) peaks is observed. This is accompanied by a concomitant red-shift (~10 cm$^{-1}$) in the CoM of both modes (see **Supplementary Note 3**). These observations are consistent with previous literature for charging of layered metal oxides used in Li-ion batteries[32] and the *ex-situ* measurements described above, and confirm that during the initial anodic scan delithiation occurs. Upon the subsequent cathodic scan, the 640 cm$^{-1}$ peak intensity increases again, however it is to an intensity lower than that of the initial α-Li$_2$IrO$_3$ phase. The overall behaviour recorded during the initial cycle is consistent with an initial oxidation/delithiation to form a phase with a chemical composition close to α-Li$_1$IrO$_3$, followed by the intercalation of hydrated potassium as the potential is dropped. After the first scan, subsequent anodic scans in the same potential range result in a dimming of the 640 cm$^{-1}$ mode by around ~40% and only a small (~5-10%) increase in the intensity of the 550 cm$^{-1}$ mode (see **Supplementary Note 3** for CoM changes). These changes in mode intensity are reversible, *i.e.* the intensity of the 640 cm$^{-1}$ (550 cm$^{-1}$) increases (decreases) during the corresponding cathodic scans/reduction. We note that following the initial cycle, the intensity of the 640 cm$^{-1}$ only begins to drop (increase) in the anodic (cathodic) scan after ~1.37 V vs RHE, indicating an oxidation event at the potential previously assigned to the onset of K$^+$ deintercalation[13]. This is the direct visualization that the active form of the catalyst, *i.e.* α-Li$_1$IrO$_3$, is regenerated electrochemically at a potential concomitant with the OER. Based on these observations we can confirm that the intensity of the 640 cm$^{-1}$ mode can be used to track reversible cation intercalation and distinguish between the various phases.

Interestingly, when switching the electrolyte to 1M LiOH, after the initial anodic scan, no drastic modification of the 640 cm$^{-1}$ (or 550 cm$^{-1}$) mode intensity is observed (**Supplementary Note 5**). In other words, no Li$^+$ intercalation occurs under these conditions due to the larger hydrodynamic radius of Li$^+$ as compared to K$^+$, in agreement with our previous findings[13]. Furthermore, for crystalline IrO$_2$ for which no bulk cation exchange was reported, two broad Raman bands around 500 cm$^{-1}$ and 700 cm$^{-1}$ are observed and cycling in the potential range of 1.1 to 1.7V vs RHE does not result in significant modification of the intensity of both bands (**Supplementary Note 5**).

Having identified vibrational signatures of cation (de)intercalation, images of the integrated intensity of the 640 $cm^{-1}$ band across a single catalyst particle (and 550 $cm^{-1}$; see **Supplementary Note 6** and **Supplementary Videos 1 and 2**) were collected using compressive Raman imaging during each step of increasing and decreasing potential, thereby allowing us to access the spatial distribution of phase fronts associated with the different cationic processes/material stoichiometries. The spectral integration window is chosen such that changes in the peak CoM do not influence the intensities observed. During the initial anodic scan, for the 640 $cm^{-1}$ mode, a front of diminishing intensity is observed to move from the surface of particles in contact with the electrolyte to the core of the particle (**Figure 3a**, top row) as the potential is increased to 1.65 V vs. RHE. The pattern of spatial motion is in-line with a so-called 'shrinking core' behaviour which has been previously observed for Li-deintercalation in other layered oxides[33,34]. During the following cathodic scan, the integrated intensity of the 640 $cm^{-1}$ mode increases from the particle edges in a pattern and intensity consistent with the intercalation of potassium cations, as discussed in **Figure 2a**. This conclusion is reinforced by noticing that in LiOH, no change is observed during the cathodic scan (**Supplementary Note 5)**.

For the second anodic cycle, a different mechanism is spotted (second row of **Figure 3a**). The core of the particle is observed to lose Raman intensity before the exterior during the anodic scan with a shell of higher intensity remaining on reaching 1.65 V vs. RHE, in an opposite manner to the 'classical' shrinking core pattern found during the first cycle. During the cathodic scan, the core is slowly replenished in potassium, and homogeneous intensity for the 640 $cm^{-1}$ vibration mode is eventually found when back to the open circuit voltage (*i.e.* 0 $mA/cm^2_{oxide}$). This process is observed across numerous cycles as shown in the bottom row of **Figure 3a**, with the shell being consistently found to be rich in potassium (*i.e.* reduced) while the core is depleted. In comparison, no such change in intensity for the 640 $cm^{-1}$ vibration mode (and the

550 cm$^{-1}$ one) was observed when performing imaging in LiOH, apart from the initial delithiation occurring during the first cycle (see **Supplementary Note 5)**, confirming that it is related to potassium (de)intercalation from the particles.

Our observation of changes throughout the particle demonstrate that the α-Li$_2$IrO$_3$ catalyst stores charges as function of potential during the OER but, unlike for amorphous or homogeneous catalysts, charges are stored in the bulk following the reversible (de)intercalation of cations. To further confirm this effect, the amount of charges stored upon OER was determined as function of applied potential by following the methods proposed by Nong *et al.*[10]. Plotting the charge accumulated as a function of the logarithm of the OER current in **Figure 3b**, a linear correlation is found in KOH, similar to that previously reported for supported IrO$_x$ nanoparticles[10]. This result contrasts with the Tafel plots, *i.e.* the overpotential as function of the logarithm of the current, obtained from the same measurements (**Figure 3c**) which show a break in slope in KOH when reaching 10 mA/cm$^2_{oxide}$, or approx. 1.6 V vs. RHE. Thus, charge stored upon cycling controls the OER current rather than potential. Unlike in KOH, no linear correlation with charge or no substantial break in Tafel slope is recorded in LiOH, reinforcing our conclusion.

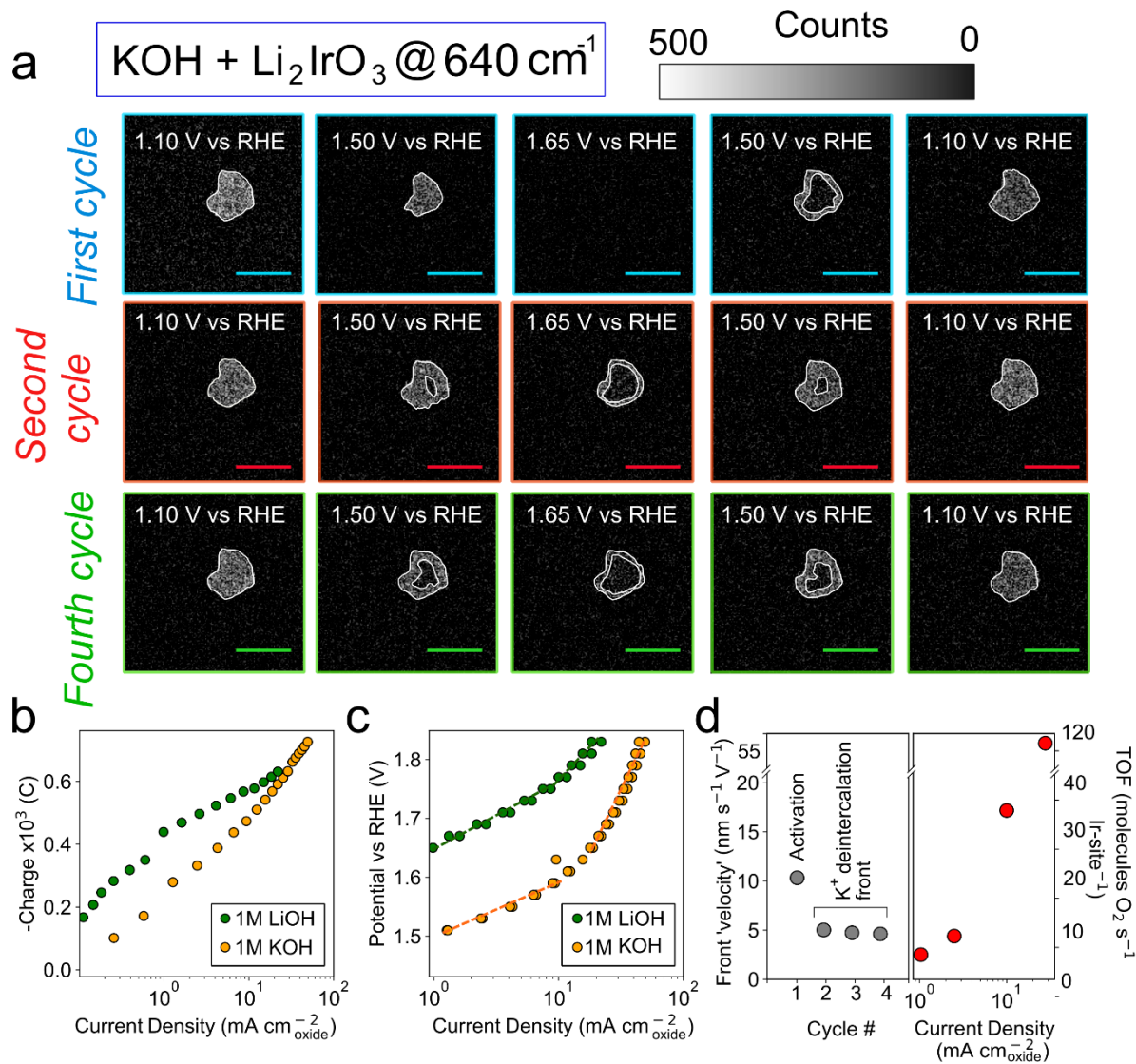

**Figure 3. a.** Raman intensity images of 640 cm$^{-1}$ modes at selected potentials during the first, second and fourth electrochemical cycles of α-Li$_2$IrO$_3$ using KOH as the electrolyte. Cycling is performed at 4.0 mV/s (see **Supplementary Videos 1-2** for other cycles and **Supplementary Videos 3-4** and **Supplementary Note 7** for repeats over more particles). Scale bar is 5 μm. **b.** Charge accumulated as a function of OER current for α-Li$_2$IrO$_3$. **c.** Tafel plot for α-Li$_2$IrO$_3$ with dotted lines indicating slope. Current density is normalized by the BET surface area of the iridium oxide catalysts. **d.** (Left, grey dots) Pseudo 'velocity' of phase front which is dominated by Li$^+$ deintercalation during initial anodic scan and K$^+$ deintercalation in scans 2 to 4. All values are extracted from the anodic scan (see **Supplementary Note 8**). (Right, red dots) Turnover frequency (TOF) for the OER on surface active sites calculated from the specific current density.

In order to understand the core-shell structure forming in the anodic scan, further analysis is required. From the *operando* Raman imaging data, a pseudo phase front 'velocity' associated with cation (de)intercalation can be extracted for our scan-rates, from the time-taken for an intensity front to propagate a set distance normalised by the voltage change (we note that strictly the term velocity refers to constant potential measurements; see **Supplementary Note 8**). First, upon initial oxidation and delithiation of $\alpha$-$Li_2IrO_3$ to $\alpha$-$Li_1IrO_3$, a phase front 'velocity' of $\approx 10 \pm 3$ nm s$^{-1}$ V$^{-1}$ is determined. Although the phase front is dominated by Li$^+$ deintercalation (see discussion above), the convolution between mechanisms means we cannot exclusively assign the 'velocity' to Li$^+$ ions, but will closely reflect it. During the subsequent cycles upon which hydrated K$^+$ cation are (de)intercalated, a front 'velocity' of $\approx 4.5 \pm 1.5$ nm s$^{-1}$ V$^{-1}$ is measured, which does not change significantly between cycles (**Figure 3d**). This latter value is extracted by examining the propagation of the intensity from the centre of the particle by ~1 µm and matches well also with the 'velocity' obtained for the reintercalation of potassium during the first cathodic scan (see **Supplementary Note 8** for further discussion).

We can compare the propagation of these phase fronts and the turnover frequency (TOF) for the OER on surface active sites, that is deduced from current density and BET surface area **(Supplementary Note 9),** as shown on the right side of **Figure 3d**. At a current density of 10 mA/cm$^2$_{oxide} reached at $\approx 1.59$ V vs RHE (**Figure 3b and c**), *i.e.* an overpotential of 360 mV, a TOF of 36 molecules of oxygen generated per second per iridium active site is calculated (considering a density of 0.043 iridium active site per Å$^2$ for (00l) facets, these facets being the ones predominantly exposed for platelets-like particles as discussed previously). At lower overpotential of 300 mV, a TOF of $\approx 9$ per active site per second is estimated, this value falling to 4 at 260 mV overpotential (*i.e.* 1 mA/cm$^2$_{oxide}). For the EC mechanism to proceed, potassium cations must intercalate into the structure to balance the charge during the OER (**Figure 1a**). Thus, for large TOF values, *i.e.* for large current densities and overpotentials, potassium cations do not have sufficient time to diffuse into the bulk of the particle to accommodate for the fast evolution of molecular oxygen (governed by the current densities). Therefore, our *operando* Raman imaging study reveals that at high current densities, molecular oxygen is predominantly evolved following a classical electrocatalytic adsorbate mechanism which dominates the electrocatalytic current of the layered iridate catalyst. At lower current densities, both the EC and the classical adsorbate mechanisms can coexist.

This quantitative analysis helps explaining the observed Raman images for which a potassium-rich periphery is observed during the anodic scan while the bulk is found depleted in potassium. During the anodic scan, K$^+$ deintercalation is driven resulting in a decrease in the intensity of the 640 cm$^{-1}$ mode in a shrinking-core type spatial pattern on the particle (*i.e.* from the edges of the particle first, alike the initial Li$^+$ deintercalation spotted during the first anodic scan). However, the oxidized particles simultaneously react chemically with the KOH electrolyte re-intercalating K$^+$ into the particle, resulting in an increase in the Raman intensity starting from the edges. The competition of these two effects of K$^+$ entering and leaving the particle will give rise to the observed spatial intensity pattern, as summarised in **Figure 4**. The above competition between K$^+$ electrochemical deintercalation and K$^+$ insertion following the EC mechanism is revealed during dynamic potential measurements, *i.e.* cyclic voltammetry. A question hence remains as to whether under *quasi*-static equilibrium similar patterns would be observed.

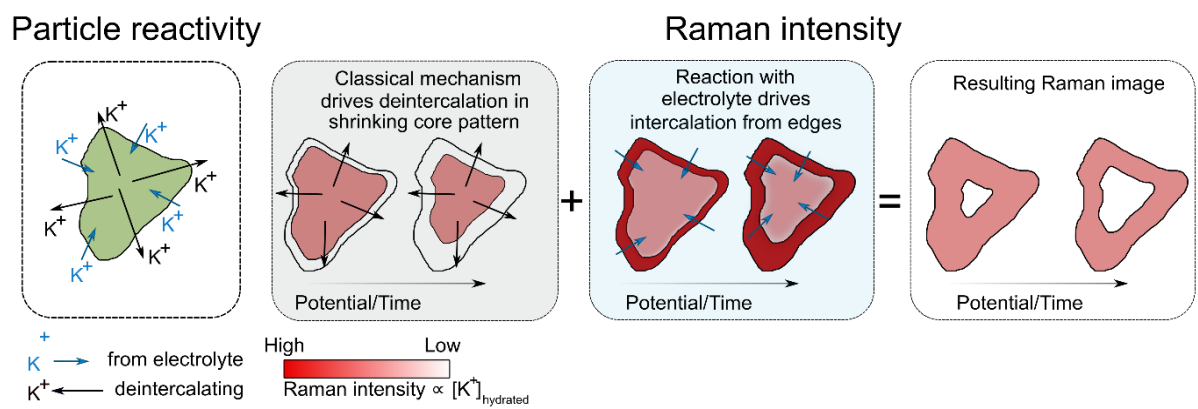

**Figure 4:** Cartoon schematic summarising how competition between bulk particle deintercalation (in a shrinking core pattern) and intercalation of K$^+$ from the electrolyte (starting from the edge of particles) gives rise to the observed spatial distribution of Raman intensity in the images after the first scan.

To verify this assertion of competing mechanisms, potentiostatic holding experiments were carried out at different potentials after two initial cycles of the catalyst (**Figure 5a** and **Supplementary Video 4-7**). Ramping up to and holding at a potential of 1.6 and 1.7 V vs RHE for 2 minutes, Raman imaging (**Figure 5b**) reveals that the full activation of the particles is not readily complete (images B and C). A potassium-rich shell is formed during the initial part of the holding, as previously observed during CV scans (**Figure 3**), before disappearing to reveal a fully oxidized particle at the end of the 2 minutes holding (image E). Hence, despite the charged particles having sufficient oxidative power to chemically evolve oxygen and insert potassium, potassium cations do not have sufficient time to diffuse into the bulk of the particle to accommodate for the large TOF at such overpotentials and the particles are eventually fully deintercalated. The reaction thus proceeds by a classical surface adsorbate mechanism on the fully charged particles. Eventually, upon relaxation (release of the potential), potassium intercalates starting from the edges (images F and G), revealing that the EC mechanism serves as a charge compensation mechanism for OER catalysts that accumulate charges. For potentials below that threshold but sufficiently large for potassium to deintercalate, *i.e.* 1.4 and 1.5 V vs. RHE, the centre of the particles is first depleted in potassium, alike at higher potentials (images B and C). However, even after an activation time of 2 minutes, the shell does not disappear and the edges remains rich in potassium owing to constant intercalation of potassium by chemical reaction with the electrolyte, revealing that the core-shell pattern reflects a *quasi*-static equilibrium (image D). We note that following the potential hold, the velocity of the intensity front corresponding to reintercalating potassium is independent of the previous holding potential and is in the order of ≈ 3.3±1 nm s$^{-1}$ (**Figure 5c**), similar to the non-voltage normalised velocity we obtain during CVs (≈ 2.7±1 nm s$^{-1}$). This finding confirms that we are imaging competing rates. At low current density, thus at low TOF for the evolving oxygen, potassium exchange and intercalation is sufficiently fast to sustain the EC mechanism, and the surface is found rich in potassium. Comparing Tafel slope analysis with potentiostatic holding experiments, a threshold of ≈1.55 V vs. RHE is confirmed, below which both the classical adsorbate and the EC mechanism can coexist.

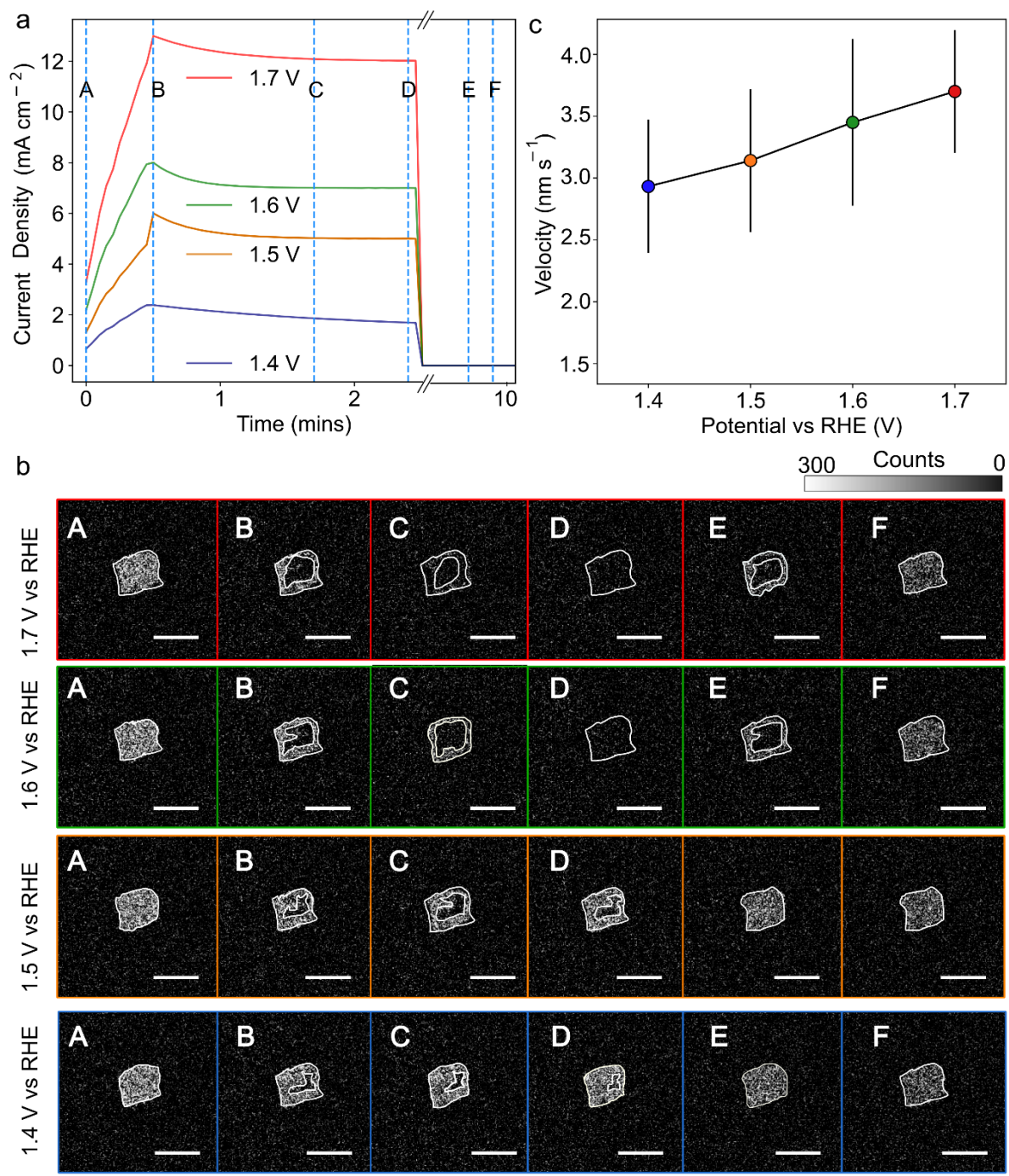

**Figure 5: a.** Ramp to potentials between 1.4 V and 1.7 V vs RHE followed by 2 min potential hold at the given potential and spontaneous release of the potential for 8 mins. **b.** Raman images integrating intensity of mode at 640 cm$^{-1}$ for single particle at set points (labelled A to F in panel a) during the potential ramp, hold and release. Scale bar is 5 μm. **c.** Velocity of phase front associated with $K^+$ ion diffusion back into structure following release of potential, extracted from spatial motion of 640 cm$^{-1}$ intensity front in regions D to F in panel.

**Conclusion**

In summary, using *operando* Raman imaging we have demonstrated the competition between two OER mechanisms for catalysts storing charges, with the current density controlling the extent to which each mechanism contributes to the evolution of oxygen. At high current density, TOF for the OER is too fast for cation exchange to occur substantially, and a classical surface adsorbate OER mechanism drives the anodic current. At lower overpotentials, the TOF becomes low enough for cations to intercalate and the EC mechanism is favoured. Our findings provide novel avenues for the design of better OER catalysts, by increasing the propagation speed of cations through the material or reducing the diffusion length to favour the EC mechanism that shows a much smaller Tafel slope compared to the classical adsorbate one.

More generally our results reveal the power of computational microscopy and *operando* Raman imaging using the compressive sensing framework. Here, the data load and analysis typically associated with hyperspectral imaging is significantly reduced, speed and sensitivity are boosted, and cost and need for expensive cameras removed[27,29,37]. The Raman spectrum of the system tackled here is relatively simple, but with appropriately designed filters[38,39] the above methods can be readily applied to more complex systems, where there may be multiple spectrally overlapped species, without losing sensitivity[29,39]. The unique liquid and solid phase chemical sensitivity and potential to be quantitative (Raman intensity is proportional to concentration), means such methods will find great use in probing other electrochemical systems, *e.g.* batteries, where intercalation mechanisms, solvent/electrolyte polarisation gradients and interfacial reactions may all be studied[40]. This is particularly in contrast to other low-cost optical techniques such as scattering/reflection microscopy[24,41] which are limited to solids and can also be challenging to interpret/make quantitative. Increasing the time-resolution of Raman (imaging) will allow faster scan rates to be explored and using optically or electronically-gated methods[18,42] it may even be possible to probe pico- to nanosecond processes such as electron-transfer and ion (de)solvation[43]. Enhancing the spatial resolution of Raman imaging with near-field[44] or super-resolution[45,46] methods will further aid answering of such questions. Automation to allow scanning over large numbers of particles and correlation with complementary techniques such as X-Ray imaging/TEM[14,47] will be key to further enhance the utility of such methods[48,49].

**References**

1. Lagadec, M. F. & Grimaud, A. Water electrolysers with closed and open electrochemical systems. *Nat. Mater. 2020 1911* **19**, 1140–1150 (2020).

2. Stamenkovic, V. R., Strmcnik, D., Lopes, P. P. & Markovic, N. M. Energy and fuels from electrochemical interfaces. *Nat. Mater. 2017 161* **16**, 57–69 (2016).

3. Hong, W. T. *et al.* Toward the rational design of non-precious transition metal oxides for oxygen electrocatalysis. *Energy Environ. Sci.* **8**, 1404–1427 (2015).

4. Fabbri, E., Habereder, A., Waltar, K., Kötz, R. & Schmidt, T. J. Developments and perspectives of oxide-based catalysts for the oxygen evolution reaction. *Catal. Sci. Technol.* **4**, 3800–3821 (2014).

5. Hong, W. T., Welsch, R. E. & Shao-Horn, Y. Descriptors of Oxygen-Evolution Activity for Oxides: A Statistical Evaluation. *J. Phys. Chem. C* **120**, 78–86 (2016).

6. She, Z. W. *et al.* Combining theory and experiment in electrocatalysis: Insights into materials design. *Science (80-. ).* **355**, (2017).

7. Man, I. C. *et al.* Universality in Oxygen Evolution Electrocatalysis on Oxide Surfaces. *ChemCatChem* **3**, 1159–1165 (2011).

8. Burke, M. S., Enman, L. J., Batchellor, A. S., Zou, S. & Boettcher, S. W. Oxygen Evolution Reaction Electrocatalysis on Transition Metal Oxides and (Oxy)hydroxides: Activity Trends and Design Principles. *Chem. Mater.* **27**, 7549–7558 (2015).

9. McCrory, C. C. L., Jung, S., Peters, J. C. & Jaramillo, T. F. Benchmarking heterogeneous electrocatalysts for the oxygen evolution reaction. *J. Am. Chem. Soc.* **135**, 16977–16987 (2013).

10. Nong, H. N. *et al.* Key role of chemistry versus bias in electrocatalytic oxygen evolution. *Nat. 2020 5877834* **587**, 408–413 (2020).

11. Pearce, P. E. *et al.* Revealing the Reactivity of the Iridium Trioxide Intermediate for the Oxygen Evolution Reaction in Acidic Media. *Chem. Mater.* **31**, 5845–5855 (2019).

12. Mefford, J. T. *et al.* Water electrolysis on La1−xSrxCoO3−δ perovskite electrocatalysts. *Nat. Commun. 2016 71* **7**, 1–11 (2016).

13. Yang, C. *et al.* Cation insertion to break the activity/stability relationship for highly active oxygen evolution reaction catalyst. *Nat. Commun. 2020 111* **11**, 1–10 (2020).

14. Mefford, J. T. *et al.* Correlative operando microscopy of oxygen evolution electrocatalysts. *Nature* **593**, (2021).

15. Saeed, K. H., Forster, M., Li, J. F., Hardwick, L. J. & Cowan, A. J. Water oxidation intermediates on iridium oxide electrodes probed by in situ electrochemical SHINERS. *Chem. Commun.* **56**, 1129–1132 (2020).

16. Li, J. F. *et al.* Shell-isolated nanoparticle-enhanced Raman spectroscopy. *Nature 2010 4647287* **464**, 392–395 (2010).

17. Zhang, H. *et al.* In situ dynamic tracking of heterogeneous nanocatalytic processes by shell-isolated nanoparticle-enhanced Raman spectroscopy. *Nat. Commun.* **8**, (2017).

18. Pasquini, C., D'Amario, L., Zaharieva, I. & Dau, H. Operando Raman spectroscopy tracks oxidation-state changes in an amorphous Co oxide material for electrocatalysis of the oxygen evolution reaction. *J. Chem. Phys.* **152**, 194202 (2020).

19. Zhu, Y. *et al.* Operando unraveling of the structural and chemical stability of P-substituted $CoSe_2$ electrocatalysts toward hydrogen and oxygen evolution reactions in alkaline electrolyte. *ACS Energy Lett.* **4**, 987–994 (2019).

20. Huang, J. *et al.* Identification of Key Reversible Intermediates in Self-Reconstructed Nickel-Based Hybrid Electrocatalysts for Oxygen Evolution. *Angew. Chemie Int. Ed.* **58**, 17458–17464 (2019).

21. Kuai, C. *et al.* Revealing the Dynamics and Roles of Iron Incorporation in Nickel Hydroxide Water Oxidation Catalysts. *J. Am. Chem. Soc.* **143**, 18519–18526 (2021).

22. Cheng, Q. *et al.* Operando and three-dimensional visualization of anion depletion and lithium growth by stimulated Raman scattering microscopy. *Nat. Commun. 2018 91* **9**, 1–10 (2018).

23. Scotté, C. *et al.* Assessment of Compressive Raman versus Hyperspectral Raman for Microcalcification Chemical Imaging. *Anal. Chem.* **90**, 7197–7203 (2018).

24. Chen, Y. *et al.* Operando video microscopy of Li plating and re-intercalation on graphite anodes during fast charging. *J. Mater. Chem. A* **9**, 23522–23536 (2021).

25. Jin, Y. *et al.* In operando plasmonic monitoring of electrochemical evolution of lithium metal. *Proc. Natl. Acad. Sci. U. S. A.* **115**, 11168–11173 (2018).

26. Lin, H. & B. de Aguiar, H. Compressive Raman microspectroscopy. in *Stimulated Raman Scattering Microscopy* (2022).

27. Cebeci, D., Mankani, B. R. & Ben-Amotz, D. Recent Trends in Compressive Raman Spectroscopy Using DMD-Based Binary Detection. *J. Imaging* **5**, 1 (2018).

28. Wilcox, D. S. *et al.* Digital compressive chemical quantitation and hyperspectral imaging. *Analyst* **138**, 4982–4990 (2013).

29. Sturm, B. *et al.* High-Sensitivity High-Speed Compressive Spectrometer for Raman Imaging. *ACS Photonics* **6**, 25 (2019).

30. Gao, J. *et al.* Breaking Long-Range Order in Iridium Oxide by Alkali Ion for Efficient Water Oxidation. *J. Am. Chem. Soc.* **141**, 3014–3023 (2019).

31. Pei, S. *et al.* Magnetic Raman continuum in single-crystalline $H_3LiIr_2O_6$. *Phys. Rev. B* **101**, 201101 (2020).

32. Flores, E., Novák, P. & Berg, E. J. In situ and Operando Raman spectroscopy of layered transition metal oxides for Li-ion battery cathodes. *Front. Energy Res.* **6**, 82 (2018).

33. Fraggedakis, D. *et al.* A scaling law to determine phase morphologies during ion intercalation.

*Energy Environ. Sci* **13**, 2142 (2020).

34. Bazant, M. Z. Theory of chemical kinetics and charge transfer based on nonequilibrium thermodynamics. *Acc. Chem. Res.* **46**, 1144–1160 (2013).

35. Geng, Z., Chien, Y. C., Lacey, M. J., Thiringer, T. & Brandell, D. Validity of solid-state Li+ diffusion coefficient estimation by electrochemical approaches for lithium-ion batteries. *Electrochim. Acta* **404**, 139727 (2022).

36. Xie, J. *et al.* Orientation dependence of Li-ion diffusion kinetics in LiCoO2 thin films prepared by RF magnetron sputtering. *Solid State Ionics* **179**, 362–370 (2008).

37. Soldevila, F., Dong, J., Tajahuerce, E., Gigan, S. & De Aguiar, H. B. Fast compressive Raman bio-imaging via matrix completion. *Optica* **6**, 341–346 (2019).

38. Mankani, B. R. *et al.* Binary Complementary Filters for Compressive Raman Spectroscopy. *Appl. Spectrosc. Vol. 72, Issue 1, pp. 69-78* **72**, 69–78 (2018).

39. Scotté, C., Galland, F., Rigneault, H., Aguiar, H. B. de & Réfrégier, P. Precision of proportion estimation with binary compressed Raman spectrum. *JOSA A* **35**, 125–134 (2018).

40. Lang, S., Yu, S. H., Feng, X., Krumov, M. R. & Abruña, H. D. Understanding the lithium–sulfur battery redox reactions via operando confocal Raman microscopy. *Nat. Commun. 2022 131* **13**, 1–11 (2022).

41. Merryweather, A. J., Schnedermann, C., Jacquet, Q., Grey, C. P. & Rao, A. Operando optical tracking of single-particle ion dynamics in batteries. *Nature* **594**, 522–528 (2021).

42. Hardwick, L. S. J. Kerr gated Raman spectroscopy of LiPF6 salt and LiPF6-based organic carbonate electrolyte for Li-ion batteries. *Phys. Chem. Chem. Phys.* **21**, 23833

43. Tarascon, J.-M. Material science as a cornerstone driving battery research. *Nat. Mater. 2022 219* **21**, 979–982 (2022).

44. Zhang, W., Fang, Z. & Zhu, X. Near-field Raman spectroscopy with aperture tips. *Chem. Rev.* **117**, 5095–5109 (2017).

45. Graefe, C. T. *et al.* Far-Field Super-Resolution Vibrational Spectroscopy. *Anal. Chem.* **91**, 8723–8731 (2019).

46. Guilbert, J. *et al.* Label-free super-resolution chemical imaging of biomedical specimens. *bioRxiv* 2021.05.14.444185 (2021).

47. Shen, T. H., Spillane, L., Peng, J., Shao-Horn, Y. & Tileli, V. Switchable wetting of oxygen-evolving oxide catalysts. *Nat. Catal. 2022 51* **5**, 30–36 (2021).

48. Godeffroy, L. *et al.* Bridging the Gap between Single Nanoparticle Imaging and Global Electrochemical Response by Correlative Microscopy Assisted By Machine Vision. *Small Methods* **2200659**, 1–12 (2022).

49. Chen, W. *et al.* Formation and impact of nanoscopic oriented phase domains in electrochemical crystalline electrodes. *Nat. Mater. 2022* 1–8 (2022).

50. McCalla, E. *et al.* Visualization of O-O peroxo-like dimers in high-capacity layered oxides for Li-ion batteries. *Science (80-. ).* **350**, 1516–1521 (2015).

51. Wang, T. & Dai, L. Background Subtraction of Raman Spectra Based on Iterative Polynomial Smoothing. *Appl. Spectrosc.* **71**, 1169–1179 (2016).

# Experimental Methods

## Preparation of the working electrodes

α-Li$_2$IrO$_3$ was synthetized according to literature[50]. IrO$_2$ was purchased from Alfa Aesar (Premion®, 99.99% metals basis, Ir 84.5% min). Electrodes were prepared by drop-casting an ink containing oxide catalyst powder on ITO-coated coverslides (400 mm², 0.15 – 0.17 mm thick, Diamond Coatings LtD) or 0.15 – 0.17 mm thick coverslides with ~10 nm of Ti evaporated atop of them (see **Supplementary Note 2**). The ink was prepared by sonicating 5 mg of catalyst in 970 μL of tetrahydrofuran (anhydrous, ≥ 99.9%, inhibitor-free, Sigma-Aldrich) for 1h. 30 μl of a Nafion D-520 dispersion (5% w/w in water and 1-propanol, ≥ 1 meq/g exchange, Alfa Aesar) were then added to the dispersion. The as-obtained dispersion was gently shook and subsequently dropcasted onto the ITO or Ti electrodes.

## Electrochemical setup

An Ag wire (0.5 mm diameter; Sigma) was attached to the conductive coverslide using epoxy glue and conductive copper tape to allow connection to the working electrode. The reference (Ag/AgCl porous frit electrode; redox.me) and counter (Pt wire; 0.5 mm thickness; Sigma) were well separated and the reference was held just above the surface of the coverslides. Micrometre screw gauges were used to achieving exact positioning. Silicone rectangular or circular shaped wells (Mcmaster Carr) with a predefined area were bonded to the sample such that the active area with electrolyte could be determined. Wells were filled with the given electrolyte (200 μL to ~1500 μL depending on the experiment) to ensure all electrodes were wet. A Gamry, Reference 600 potentiostat was used to apply all potentials (see **Supplementary Note 10** for further details).

## Raman Microspectroscopy

The microscope was a standard layout of an epi-detected Raman microscope. A pump laser beam (wavelength = 532 nm, Coherent Mira) was spectrally cleaned up by a bandpass filter (FLH05532-4, Thorlabs), and its beam width was expanded to 7.2 mm before entering a home-built inverted microscope. Additional waveplates (half-waveplate and quarter-waveplate for 532 nm, Foctek Photonics) precompensated the ellipticity introduced by the dichroic filter (F38-532_T1, AHF) and also generated circularly polarized light. We used high numerical aperture (NA) oil-immersion objectives (Nikon 60X/1.4NA oil) to ensure high-resolution imaging and increase collection efficiency. The maximum pump power before the objective was ~50 mW, a power level that ensured no degradation of samples. The samples were scanned with galvanometric mirrors (Thorlabs). The Raman inelastic backscattered light was collected by the same objective and focused with the microscope tube lens either onto the slit of the conventional spectrometer (Andor, Shamrock 303i, grating 300 l/mm; the slit also acts an effective pinhole for confocal detection) or a 30 mm line slit (Thorlabs) which is at the entrance of the home-built compressive spectrometer whose specifications are detailed in Sturm et al.[29]. A notch filter blocked residual pump light (NF533-17, Thorlabs) before guiding the signal to the spectrometers. The conventional spectrometer is equipped with a high-sensitivity charge-coupled camera (Andor, iXon 897). All images presented with the compressive spectrometer were taken with integration times/pixel in the 20 to 50 μs range. Imaging is performed on flat particles which are identified as being 'flat' if the entire particle appears to be in focus under bright field illumination, *i.e.* within our axial resolution of ~600 nm the surface of the particle is in the same imaging plane. Recording of data was performed by a custom Matlab program with external synchronisation to the potentiostat for control of the potential whilst recording Raman hyperspectra/images. For spectrally resolved data background subtraction was performed using a modified iterative polynomial smoothing method[51]. In **Supplementary Note 11** we discuss the 'low-cost' nature of our approach and its comparison to other benchtop characterisation methods.

**Acknowledgements**

R.P thanks Clare College, University of Cambridge for funding the work *via* a Junior Research Fellowship, Scott Keene (Cambridge) for critical reading and feedback on the manuscript, Thierry Barisien, Erwan Dandeu and Loic Becerra (INSP) for preparation of Ti coverslides and Prof. Ulrich F. Keyser (Cambridge) for loan of a potentiostat and laser. F.D acknowledges the École normale supérieure Paris-Saclay for his PhD scholarship. D.R. and A.W.C. acknowledge support from the ANR project ACCEPT (Grant no. ANR-19-CE24-0028). D.R. thanks L. Maschio for fruitful discussions on the simulations of cations with CRYSTAL17. H.B.A. acknowledges support from Sorbonne Université (Emergence – FANCIER). A.G. thanks the French National Research Agency for its support through the Labex STORE-EX project (ANR-10LABX-76-01). This work was granted access to the HPC resources of IDRIS, CINES and TGCC under the allocation 2022-A0120912417 made by GENCI.

**Author contributions**

R.P performed Raman experiments and analysed the data under the guidance of H.B.A.. F.D. prepared the samples, performed calculations on the electrochemical data and interpreted the results. D.R. performed and interpreted the DFT calculations under the supervision of A.W.C.. S.G. provided feedback on imaging data. A.G. conceived the idea, designed experiments and supervised the project. All authors contributed to writing of the manuscript.